\documentclass{article}%
\usepackage{amssymb}
\usepackage{amsfonts}
\usepackage{amsmath}
\usepackage{graphicx}%
\providecommand{\U}[1]{\protect\rule{.1in}{.1in}}

\begin{document}

\title{A general statistical approach to quantum algorithms in a circuit model based
on the expectation and standard deviation\ of each gate separately}
\author{Tomer Shushi\\Center for Quantum Science and Technology\\\& Department of Business Administration,\\Guilford Glazer Faculty of Business and Management,\\Ben-Gurion University of the Negev, Beer-Sheva, Israel}
\maketitle

\begin{abstract}
Recently, there has been a growing literature exploring the generalization of
quantum algorithms, such that different quantum algorithms are special
examples of a more fundamental structure. In this short paper, we provide a
general approach to describe quantum algorithms as a quantum state with
amplitudes that are constructed from the expected value and standard deviation
of each quantum gate or a sub-sequence of gates in the algorithm. The proposed
statistical-based description relies on the celebrated Aharonov-Vaidman
identity. We present a more fundamental identity that, unlike the previous
one, allows us to switch the basis of the states into a desired form.


\end{abstract}

\section{Introduction}

Quantum algorithms are a cornerstone of quantum computing, exploiting the
unique principles of quantum mechanics---such as superposition and
entanglement---to solve problems more efficiently than classical algorithms.
Famous quantum algorithms, like Shor's algorithm for factoring and Grover's
search algorithm, offer exponential and quadratic speedups, respectively. As
quantum computing evolves, exploring more fundamental structures that unify
different quantum algorithms becomes increasingly important, potentially
revealing deeper insights into quantum computation. One promising approach is
through amplitude estimation algorithms [1-3], which generalize Grover's
search by efficiently estimating the probability amplitudes of quantum states.
This class of algorithms plays a key role in quantum machine learning,
optimization, and financial modeling, and it represents a foundational tool in
quantum computing by reducing the complexity of tasks such as Monte Carlo
simulations. Quantum signal processing (see [4-5]) is a general framework that
is based on applying polynomial functions to eigenvalues of unitary matrices
using controlled rotations. Another important framework is the quantum
singular value transformation that builds up a unified method in which
different quantum algorithms are deduced from a more basic structure [6]. In
[7], the authors have shown how such a method embeds major quantum algorithms
such as amplitude amplification, quantum linear systems problem, and quantum
simulation. Quantum block encoding is a method that allows to represent a
given (matrix) operator as a submatrix embedded within a larger unitary
matrix, allowing a more flexible way to build up quantum circuits (see, e.g.,
[8,9]) and currently, the most efficient way to follow this method is by using
the quantum singular value transformation and quantum signal processing (see,
[10]). In [11], the authors proposed a unified framework for quantum
classification, followed by trainable quantum circuits. In this short paper,
we present a general framework for translating the operations on the qubits in a circuit model 
into quantum states with amplitudes that depend on the expectation and
standard deviation of the gates in the algorithm.

\section{\bigskip Results}

Since its appearance about 35 years ago, the Aharonov-Vaidman (AV) identity
has been used and well-explored as a fundamental relation in quantum mechanics
[12-17]. This fundamental identity relates the action of an observable
operator on a quantum state to its expected value and a projection onto a
subspace orthogonal to the state. Mathematically, the AV identity is
a\ universal formula for a Hermitian operator $A$ acting on a quantum state
$\left\vert \psi\right\rangle ,$%
\begin{equation}
A\left\vert \psi\right\rangle =\left\langle A\right\rangle ^{\psi}\left\vert
\psi\right\rangle +\Delta_{\psi}A\left\vert \psi_{\bot}\right\rangle ,
\label{AV1}%
\end{equation}
where\ $\left\vert \psi_{\bot}\right\rangle $ is a quantum state that is
orthogonal to $\left\vert \psi\right\rangle ,$ $\left\langle A\right\rangle
^{\psi}=\left\langle \psi\right\vert A\left\vert \psi\right\rangle $ is the
expectation of $A$ with respect to the state vector, and $\Delta_{\psi
}A:=\sqrt{\left\langle \psi\right\vert \left(  A-\left\langle A\right\rangle
\right)  ^{2}\left\vert \psi\right\rangle }$ is the associated quantum
uncertainty of $A.$ We see that the operation of $A$ on the quantum state
naturally yields to the statistical measures of expectation and standard
deviation, followed by the key role that probability plays in quantum mechanics.

The AV\ identity (\ref{AV1}) has been used for various aspects of quantum
mechanics, such as weak measurements and weak values (see, e.g., [15]), and
recently, it has been shown that this identity allows one to deduce the Born
rule from standard quantum mechanics by postulating that a property of quantum
particles remains fixed for a sufficiently short time [17]. The AV\ identity
is at the heart of the proposed results.

We introduce a more general identity that, unlike the AV\ identity, leads to a
quantum state on a desired basis. But, before that, we provide a short
introduction to the concept of \textit{weak measurements} and \textit{weak
values}.

The two-state vector formalism (TSVF) provides a time-symmetric formulation of
quantum mechanics, where quantum systems are described by initial and final
boundary conditions of the quantum system. It is shown that this framework
opens the door for unique measurements that do not collapse the wavefunction,
known as weak measurements, where the outcomes of such measurements are known
as weak values [18-22]. Let $\left\vert \psi\right\rangle $ be our initial
(pre-selected) state of the quantum system, and let the final (post-selected)
state be $\left\langle \phi\right\vert $, which is not an orthogonal state to
$\left\vert \psi\right\rangle $. Then, the weak value of an observable $A$ is
defined as%
\begin{equation}
\left\langle A\right\rangle _{w}^{\phi,\psi}:=\frac{\left\langle \phi
|A|\psi\right\rangle }{\left\langle \phi|\psi\right\rangle },\text{
\ }\left\langle \phi|\psi\right\rangle \neq0. \label{Weak1}%
\end{equation}
Unlike strong measurements, weak values can take on values outside the range
of the eigenvalues of the measured observable $A.$ Moreover, unlike the
standard quantum expectation $\left\langle A\right\rangle ^{\psi},$ the weak
value $\left\langle A\right\rangle _{w}^{\phi,\psi}$ is, in general, complex-valued.

In the following, we provide a more fundamental identity for the operation of
an operator on the quantum state into a superposition of states in a chosen
basis. To obtain the proposed identity, we consider, again, a pair of quantum
states $\left\vert \psi\right\rangle $ and $\left\vert \phi\right\rangle $
that are not orthogonal. We can define an orthogonal state to $\left\vert
\phi\right\rangle $ which we term as $\left\vert \phi_{\perp}\right\rangle $
such that%
\begin{equation}
\left\vert \phi_{\perp}\right\rangle =\frac{1}{C}R_{\phi}A\left\vert
\psi\right\rangle ,\label{S_phi1}%
\end{equation}
where $C\neq0$ is the normalizing constant such that $\left\langle \phi
_{\perp}|\phi_{\perp}\right\rangle =1,$ and
\[
R_{\phi}:=qI-K\left\vert \phi\right\rangle \left\langle \phi\right\vert
\]
is an orthogonal projection operator, where $q\in%
\mathbb{R}
_{>0}$ is some constant and $K$ is an operator that satisfies
\begin{equation}
\left\langle K\right\rangle ^{\phi}=q.\label{Kq1}%
\end{equation}
Follow such a construction, it is then evident that $\left\vert \phi
\right\rangle $ and $\left\vert \phi_{\perp}\right\rangle $ are orthogonal to
each other%
\begin{equation}
\left\langle \phi|\phi_{\perp}\right\rangle =\frac{1}{C}\left(  q\cdot
\left\langle \phi\right\vert A\left\vert \psi\right\rangle -\left\langle
K\right\rangle ^{\phi}\cdot\left\langle \phi\right\vert A\left\vert
\psi\right\rangle \right)  =0.\label{ortho_q1}%
\end{equation}
The normalization condition $\left\langle \phi_{\perp}|\phi_{\perp
}\right\rangle =1$ leads to the calculation of the normalizing constant, which
takes the form
\begin{equation}
C=q\mu_{R_{\phi}A}^{\psi}\Delta_{\psi}\left(  R_{\phi}A\right)  \label{CC1}%
\end{equation}
where $\mu_{R_{\phi}A}^{\psi}:=\sqrt{1+\left(  \left\langle R_{\phi
}A\right\rangle ^{\psi}\right)  ^{2}/\Delta_{\psi}\left(  R_{\phi}A\right)
^{2}}/\left\langle K\right\rangle ^{\phi}.$ We note that in the case
\thinspace$q=1$ and $K=I,$ when considering that the state $\left\vert
\phi\right\rangle $ is part from a full basis with $\sum_{j=1}\left\vert
\phi_{j}\right\rangle \left\langle \phi_{j}\right\vert =I$ such that
$\left\vert \phi\right\rangle $ is taken to be $\left\vert \phi_{1}%
\right\rangle ,$ $C$ takes the form $C=q\sqrt{\sum_{j=2}\left\langle \psi
|\phi_{j}\right\rangle \left\langle \phi_{j}|\psi\right\rangle \cdot
\left\langle A^{\dag}\right\rangle _{w}^{\psi,\phi_{j}}\cdot\left\langle
A\right\rangle _{w}^{\phi_{j},\psi}}.$ We further define $\kappa:=\left\langle
\phi|\psi\right\rangle /\left\langle K\right\rangle ^{\phi}\neq0.$

Substituting (\ref{Weak1}) and (\ref{CC1}) into (\ref{S_phi1}), after some
calculations, we obtain the desired extension of the AV identity%
\begin{equation}
A\left\vert \psi\right\rangle =\left\langle \kappa A\right\rangle _{w}%
^{\phi,\psi}K\left\vert \phi\right\rangle +\mu_{R_{\phi}A}^{\psi}\Delta_{\psi
}\left(  R_{\phi}A\right)  \left\vert \phi_{\perp}\right\rangle .
\label{New_inde1}%
\end{equation}
We see that, unlike the AV\ identity, we have decomposed the operation
$A\left\vert \psi\right\rangle $ into a superposition of an operation $K$
acting on a state $\left\vert \phi\right\rangle $ and its orthogonal state.
For instance, if we have a system of $N$ qubits with the initial state
$\left\vert 0\right\rangle ^{\otimes N}$ and a gate $G_{0},$ we can choose a
desired basis, e.g.,
\begin{equation}
\left\vert \phi\right\rangle =\left\vert +\right\rangle ^{\otimes N}=\frac
{1}{2^{N/2}}%
{\textstyle\prod\limits_{i=1}^{N}}
\left(  \left\vert 0\right\rangle _{i}+\left\vert 1\right\rangle _{i}\right)
, \label{eq_basis}%
\end{equation}
and an operation (a gate)\ acting on it, denoted by $G_{+},$ to obtain the
statistical-based translation mechanism of the quantum algorithm%
\begin{equation}
G_{0}\left\vert 0\right\rangle ^{\otimes N}=\left\langle \kappa G_{0}%
\right\rangle _{w}^{+,0}\cdot G_{+}\left\vert +\right\rangle ^{\otimes N}%
+\mu_{R_{\phi}G_{0}}^{0}\Delta_{0}\left(  R_{\phi}G_{0}\right)  \cdot
\left\vert +_{\perp}\right\rangle , \label{G011}%
\end{equation}
where $\left\vert +_{\perp}\right\rangle $ is a state that is orthogonal state
to $\left\vert +\right\rangle ^{\otimes N}.$

The following Figure illustrates the proposed identity compared with the AV identity.

\includegraphics[scale=0.5]{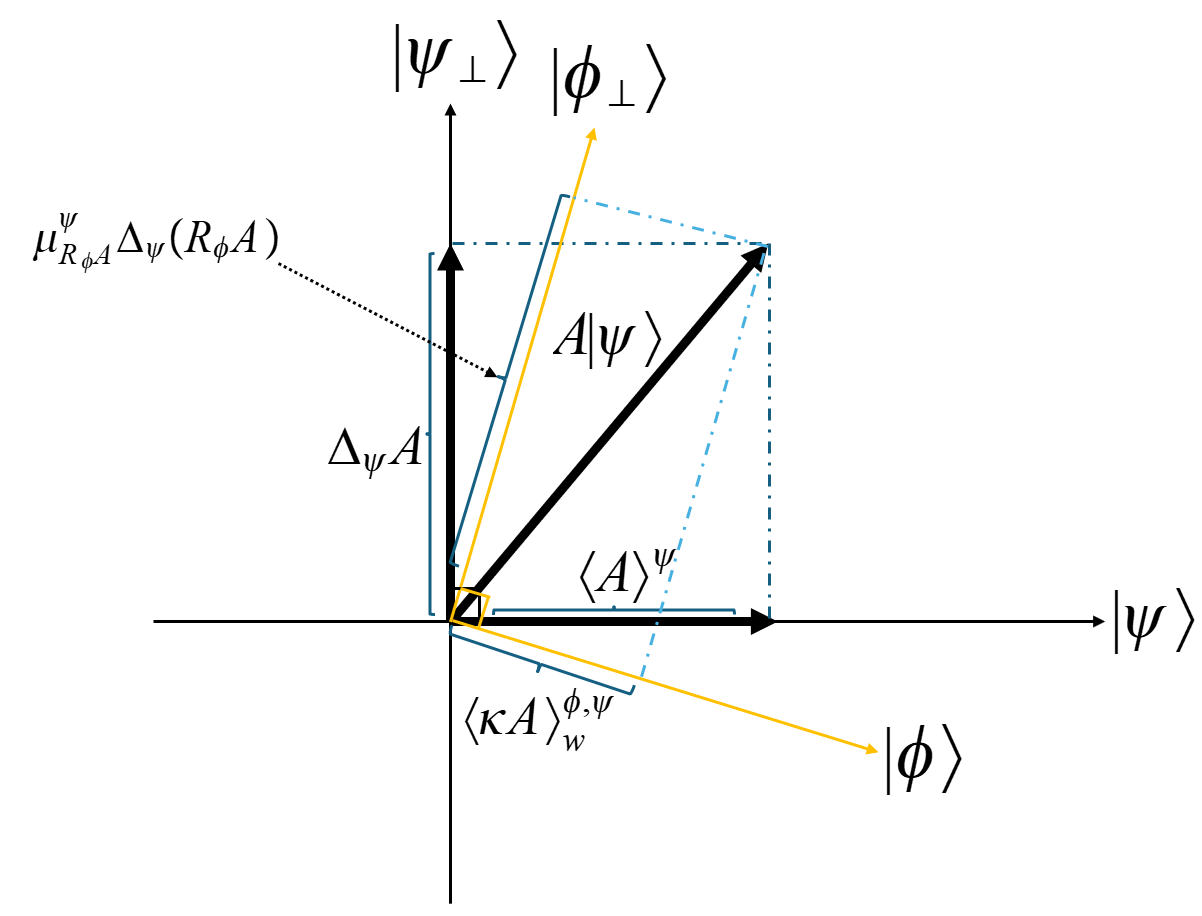}

Figure 1. A geometric illustration of the proposed identity for $K\equiv I,$
compared to the AV identity for real-valued (scaled) weak value $\left\langle
\kappa A\right\rangle _{w}^{\phi,\psi}.$

\bigskip

As illustrated in Figure 1, the operation of $A$ acting on $\left\vert
\psi\right\rangle $ can be geometrically described as a vector within the
orthogonal axes constructed from $\left\vert \phi\right\rangle $ and
$\left\vert \phi_{\bot}\right\rangle $ where $\left\vert \phi\right\rangle $
can serve as a new basis for the qubits' state. In the special case where
$\left\vert \phi\right\rangle :=\left\vert \psi\right\rangle ,$ we reconstruct
the AV identity with a similar geometry shown in [23].

Every quantum algorithm in the circuit model can be described as a sequence of
quantum gates
\begin{equation}
G_{M},G_{M-1},...,G_{0} \label{Gates1}%
\end{equation}
acting on an initial quantum state $\left\vert 0\right\rangle ^{\otimes N}$,
which we take as the zero state of $N$ qubits on the computational basis.
Then, we can define our quantum algorithm through the process%
\begin{equation}
U_{G}\left\vert 0\right\rangle ^{\otimes N}=G_{M}G_{M-1}...G_{0}\left\vert
0\right\rangle ^{\otimes N}. \label{Q_seq1}%
\end{equation}
The basic observation of the proposed result is that we can operate the
AV\ identity and its extension (\ref{New_inde1}) on the entire sequence of
operations $G_{M}G_{M-1}...G_{0}$ or sub-operations%
\begin{equation}
\mathcal{G}_{j;p}:=G_{j+p}G_{j+p-1}...G_{j},\text{ }j=0,1,...,M,\text{
}p=0,1,2,...,M-j. \label{sub_seq1}%
\end{equation}
By applying the AV identity on $\mathcal{G}_{j;p},$ we have%
\begin{equation}
\mathcal{G}_{j;p}\left\vert \psi\right\rangle =\left\langle \mathcal{G}%
_{j;p}\right\rangle ^{\psi}\left\vert \psi\right\rangle +\Delta_{\psi
}\mathcal{G}_{j;p}\left\vert \psi_{\perp}\right\rangle , \label{Formula1}%
\end{equation}
where $\left\vert \psi_{\perp}\right\rangle $ is an orthogonal state (of the
$N$ qubits) to $\left\vert \psi\right\rangle .$ When applying (\ref{Formula1})
on the entire sequence, we have a superposition of only two states, the state
$\left\vert 0\right\rangle ^{\otimes N}$ and its associated orthogonal state
$\left\vert 0_{\perp}\right\rangle ,$ while their amplitudes $\left\langle
\mathcal{G}_{j;p}\right\rangle ^{\psi=0}=\left\langle 0\left\vert ^{\otimes
N}\mathcal{G}_{j;p}\right\vert 0\right\rangle ^{\otimes N}$ and $\Delta
_{\psi=0}\mathcal{G}_{j;p},$ respectively, are based on the statistical
measures of the sequence of gates. We can simplify the calculation of the
amplitudes by applying (\ref{Formula1}) on sub-operations as given in
(\ref{sub_seq1}). Then, the amplitudes of the desired state consist of
combinations of expectations and standard deviations of the sub-sequences,
where in case of $p=0,$ the amplitudes of the final state are combinations of
expectations and standard deviations of each of the operations separately,
i.e., the amplitudes are functions of the pairs
\begin{equation}
\left\{  \left\langle G_{j}\right\rangle ^{\psi},\Delta_{\psi}G_{j}\right\}
_{j=0}^{M},\text{ } \label{G_pairs}%
\end{equation}
for associated states $\psi$.

Then, the quantum algorithm $U_{G}\left\vert 0\right\rangle ^{\otimes N}%
\ $leads to a desired state of the form%
\begin{equation}
U_{G}:\left\vert 0\right\rangle ^{\otimes N}\longmapsto\sum_{x\in\left\{
0,1\right\}  ^{N}}c_{x}\left(  \left\{  \left\langle G_{j}\right\rangle
^{\Psi},\Delta_{\Psi}G_{j}\right\}  _{j=0}^{M}\right)  \left\vert
x\right\rangle \label{U_G}%
\end{equation}
where $\left\vert x\right\rangle $ are states in the computational basis of
the $N$ qubits.

As an illustrative example, let us consider the case of a pair of qubits and
four quantum gates, where the first gate is the NOT gate acting on the second
qubit, $G_{0}=X_{\left[  2\right]  },$ in which the Deutsch algorithm is a
special case. Then, we have
\begin{align}
G_{3}G_{2}G_{1}G_{0}\left\vert 0\right\rangle ^{\otimes2}  &  =G_{3}G_{2}%
G_{1}\left\vert 01\right\rangle =\left\langle G_{1}\right\rangle ^{01}\cdot
G_{3}G_{2}\left\vert 01\right\rangle +\left(  \Delta_{01}G_{1}\right)  \cdot
G_{3}G_{2}\left\vert a_{1\bot}b_{1\bot}\right\rangle \label{Stage1}\\
&  =\left\langle G_{1}\right\rangle ^{01}\cdot\left(  \left\langle
G_{2}\right\rangle ^{01}\cdot G_{3}\left\vert 01\right\rangle +\left(
\Delta_{01}G_{2}\right)  G_{3}\left\vert a_{2\bot}b_{2\bot}\right\rangle
\right) \nonumber\\
&  +\Delta_{01}G_{1}\cdot\left(  \left\langle G_{2}\right\rangle ^{a_{1\bot
}b_{1\bot}}G_{3}\left\vert a_{1\bot}b_{1\bot}\right\rangle +\left(
\Delta_{a_{1\bot}b_{1\bot}}G_{2}\right)  G_{3}\left\vert a_{3\bot}b_{3\bot
}\right\rangle \right)  ,\nonumber
\end{align}
applying, again, the AV\ identity, and after some algebraic calculations, we
conclude that%
\begin{equation}
G_{3}G_{2}G_{1}G_{0}\left\vert 0\right\rangle ^{\otimes2}=c_{0}\left\vert
01\right\rangle +\sum_{p=1}^{7}c_{p}\left\vert a_{p\bot}b_{p\bot}\right\rangle
. \label{Stage2}%
\end{equation}
Here%
\begin{align}
c_{0}  &  =\left\langle G_{1}\right\rangle ^{01}\left\langle G_{2}%
\right\rangle ^{01}\left\langle G_{3}\right\rangle ^{01},c_{1}=\left(
\Delta_{01}G_{1}\right)  \left\langle G_{2}\right\rangle ^{a_{1\bot}b_{1\bot}%
}\left\langle G_{3}\right\rangle ^{a_{1\bot}b_{1\bot}},\label{c17}\\
c_{2}  &  =\left\langle G_{1}\right\rangle ^{01}\left(  \Delta_{01}%
G_{2}\right)  \left\langle G_{3}\right\rangle ^{a_{2\bot}b_{2\bot}}%
,c_{3}=\left(  \Delta_{01}G_{1}\right)  \left(  \Delta_{a_{1\bot}b_{1\bot}%
}G_{2}\right)  \left\langle G_{3}\right\rangle ^{a_{3\bot}b_{3\bot}%
}\nonumber\\
c_{4}  &  =\left\langle G_{1}\right\rangle ^{01}\left\langle G_{2}%
\right\rangle ^{01}\Delta_{01}G_{3},c_{5}=\left\langle G_{1}\right\rangle
^{01}\left(  \Delta_{01}G_{2}\right)  \Delta_{a_{2\bot}b_{2\bot}}%
G_{3}\left\vert a_{5\bot}b_{5\bot}\right\rangle ,\nonumber\\
c_{6}  &  =\left(  \Delta_{01}G_{1}\right)  \left\langle G_{2}\right\rangle
^{a_{1\bot}b_{1\bot}}\Delta_{a_{1\bot}b_{1\bot}}G_{3},c_{7}=\left(
\Delta_{01}G_{1}\right)  \left(  \Delta_{a_{1\bot}b_{1\bot}}G_{2}\right)
\Delta_{a_{3\bot}b_{3\bot}}G_{3}.\nonumber
\end{align}
We note that, in general, it is possible that $\left\vert a_{p\bot}b_{p\bot
}\right\rangle =\left\vert a_{p^{\prime}\bot}b_{p^{\prime}\bot}\right\rangle $
for $p\neq p^{\prime}.$ We have thus translated the quantum algorithm into a
desired state that is constructed from the expectation and standard deviation
of each quantum gate separately.

We can find the orthogonal state $\left\vert a_{p\bot}b_{p\bot}\right\rangle $
via $\left\vert a_{\left(  p-1\right)  \bot}b_{\left(  p-1\right)  \bot
}\right\rangle $ using the formula%
\begin{equation}
\left\vert a_{p\bot}b_{p\bot}\right\rangle =\frac{1}{C_{p}}\left(
I-\left\vert a_{\left(  p-1\right)  \bot}b_{\left(  p-1\right)  \bot
}\right\rangle \left\langle a_{\left(  p-1\right)  \bot}b_{\left(  p-1\right)
\bot}\right\vert \right)  G_{j}\left\vert a_{\left(  p-1\right)  \bot
}b_{\left(  p-1\right)  \bot}\right\rangle , \label{relations1}%
\end{equation}
where $\left\vert a_{\left(  -1\right)  \bot}b_{\left(  -1\right)  \bot
}\right\rangle :=\left\vert 01\right\rangle .$ In the case of Deutsch
algorithm, we have a pair of qubits and the quantum gates $G_{0}=X_{\left[
2\right]  },G_{1}=H_{\left[  2\right]  }H_{\left[  1\right]  },G_{2}%
=U_{f},G_{3}=H_{\left[  1\right]  }.$\ By measuring the averages and standard
deviations of each of the quantum gates separately, we are able to
compute\ all of the components of the amplitudes $c_{j}.$

The Grover operator, a key sub-routine in quantum algorithms such as Grover's
quantum search algorithm [24]. This iterative transformation amplifies the
probability amplitude of the desired solution in an unstructured search space,
providing a quadratic speedup over classical algorithms. The Grover operator
takes the form $\mathcal{Q}=\mathcal{AS}_{0}\mathcal{A}^{\dag}\mathcal{S}_{f}$
where $\mathcal{S}_{f}$ is called the phase oracle, $\mathcal{S}_{0}$ is the
zero phase-shift (zero reflection) operator, and $\mathcal{A}$ is the state
preparation, which in the Grover algorithm consists of the Hadramrd gates,
$\mathcal{A=}H^{\otimes N}.$

We can decompose the computation of the Grover operator\ by applying the AV
identity of the quantum algorithm on the sub-routines $\mathcal{G}%
_{j;p}^{\left[  1\right]  }=\mathcal{A}^{\dag}\mathcal{S}_{f}$ and
$\mathcal{G}_{j;p}^{\left[  2\right]  }=\mathcal{AS}_{0}.$ Starting with some
quantum state $\left\vert \psi\right\rangle $ of the $N$ qubits, we can
decompose the sub-routine $\mathcal{Q}\ $into a computation of each part of
the sub-routine into%
\begin{equation}
\mathcal{G}_{j;p}^{\left[  2\right]  }\mathcal{G}_{j;p}^{\left[  1\right]
}\left\vert \psi\right\rangle =c_{0}\left\vert \psi\right\rangle
+c_{1}\left\vert \psi_{1\bot}\right\rangle +c_{2}\left\vert \psi_{2\bot
}\right\rangle +c_{3}\left\vert \psi_{3\bot}\right\rangle
\label{desiredState01}%
\end{equation}
where%
\begin{align}
c_{0}  &  =\left\langle \mathcal{G}_{j;p}^{\left[  1\right]  }\right\rangle
^{\psi}\left\langle \mathcal{G}_{j;p}^{\left[  2\right]  }\right\rangle
^{\psi},c_{1}=\left(  \Delta_{\psi}\mathcal{G}_{j;p}^{\left[  1\right]
}\right)  \left\langle \mathcal{G}_{j;p}^{\left[  2\right]  }\right\rangle
^{\psi_{1\bot}}\label{cc1}\\
c_{2}  &  =\left\langle \mathcal{G}_{j;p}^{\left[  1\right]  }\right\rangle
^{\psi}\Delta_{\psi}\mathcal{G}_{j;p}^{\left[  2\right]  },c_{3}=\left(
\Delta_{\psi}\mathcal{G}_{j;p}^{\left[  1\right]  }\right)  \Delta
_{\psi_{1\bot}}\mathcal{G}_{j;p}^{\left[  2\right]  }.\nonumber
\end{align}
Using the proposed identity (\ref{New_inde1}), we can derive a similar
decomposition but on a computational basis and with respect to the zero state
$\left\vert 0\right\rangle ^{\otimes N},$%
\begin{equation}
\mathcal{G}_{j;p}^{\left[  2\right]  }\mathcal{G}_{j;p}^{\left[  1\right]
}\left\vert \psi\right\rangle =c_{1}^{\prime}\left\vert 0\right\rangle
^{\otimes N}+c_{2}^{\prime}\left\vert 0_{\bot}\right\rangle +c_{3}^{\prime
}\left\vert 0_{\left(  2\right)  \bot}\right\rangle +c_{4}^{\prime}\left\vert
0_{\left(  3\right)  \bot}\right\rangle , \label{G2G111}%
\end{equation}
for $K=I.$ By applying the zero state in the computational basis, $\left\vert
\phi\right\rangle =\left\vert 0\right\rangle ^{\otimes N}$, we get that all
other states, $\left\vert 0_{\left(  p\right)  \bot}\right\rangle $, are also
in the same computational basis. So, we needed to use only the proposed
identity and then apply standard AV identity.

Quantum phase estimation is a fundamental sub-routine in quantum computing
used to estimate the eigenvalue, or phase, associated with an eigenstate of a
unitary operator (see, [25-27]). By preparing a quantum state and applying
controlled unitary operations, the algorithm leverages the power of quantum
interference to extract phase information with high precision. This process is
essential for many quantum algorithms, such as Shor's algorithm for factoring
large numbers, quantum simulations of physical systems, and solving eigenvalue
problems. Suppose we have a quantum system of $N$ qubits starting at the state
$\left\vert 0\right\rangle ^{\otimes N}$ and a quantum state $\left\vert
v\right\rangle $ constructed from $N^{\prime}$ qubits that is an eigenstate of
a unitary operator $U$ such that $U\left\vert v\right\rangle =e^{i\phi
}\left\vert v\right\rangle $ for a (real-valued) scalar $\phi.$ Then, the
(sub-routine) quantum algorithm is given by $U_{G}\left\vert 0\right\rangle
^{\otimes N}\left\vert v\right\rangle $ where $U_{G}=G_{M}...G_{0}$ where%
\begin{equation}
G_{0}=H^{\otimes N},G_{1}=C_{1}U,G_{2}=C_{2}U^{2},...,G_{N}=C_{N}U^{2^{N-1}%
},G_{M=N+1}=IQFT\label{Gates0123}%
\end{equation}
and so that $U_{G}\left\vert 0\right\rangle ^{\otimes N}\left\vert
v\right\rangle =G_{M}...G_{1}\left\vert +\right\rangle ^{\otimes N}\left\vert
v\right\rangle $ for plus states $\left\vert +\right\rangle _{i}=\frac
{1}{\sqrt{2}}\left(  \left\vert 0\right\rangle _{i}+\left\vert 1\right\rangle
_{i}\right)  .$ We can decompose the process into sub-routines%
\begin{align}
\mathcal{G}_{1;N-1} &  :=G_{M-1}...G_{1},\label{Partition1}\\
\mathcal{G}_{N+1;0} &  =IQFT,\nonumber
\end{align}
to obtain the desired statistical-based decomposition%
\begin{equation}
U_{G}\left\vert 0\right\rangle ^{\otimes N}\left\vert v\right\rangle
=c_{0}\left\vert \Psi\right\rangle +c_{1}\left\vert \Psi_{1\bot}\right\rangle
+c_{2}\left\vert \Psi_{2\bot}\right\rangle +c_{3}\left\vert \Psi_{3\bot
}\right\rangle \label{Final11}%
\end{equation}
where $\left\vert \Psi\right\rangle =\left\vert +\right\rangle ^{\otimes
N}\left\vert v\right\rangle $, with coefficients that depend on the
expectation of standard deviations of the sub-routines%
\begin{align}
c_{1} &  =\left\langle \mathcal{G}_{1;N-1}\right\rangle ^{\left\vert
+\right\rangle ^{\otimes N}\left\vert v\right\rangle }\left\langle
IQFT\right\rangle ^{\left\vert +\right\rangle ^{\otimes N}\left\vert
v\right\rangle },\label{c1231}\\
c_{1} &  =\left(  \Delta_{\left\vert +\right\rangle ^{\otimes N}\left\vert
v\right\rangle }\mathcal{G}_{1;N-1}\right)  \left\langle IQFT\right\rangle
^{\left\vert \Psi_{\bot}\right\rangle },\nonumber\\
c_{2} &  =\left\langle \mathcal{G}_{1;N-1}\right\rangle ^{\left\vert
+\right\rangle ^{\otimes N}\left\vert v\right\rangle }\left(  \Delta
_{\left\vert +\right\rangle ^{\otimes N}\left\vert v\right\rangle
}IQFT\right)  ,\nonumber\\
c_{3} &  =\left(  \Delta_{\left\vert +\right\rangle ^{\otimes N}\left\vert
v\right\rangle }\mathcal{G}_{1;N-1}\right)  \left(  \Delta_{\left\vert
\Psi_{\bot}\right\rangle }IQFT\right)  \nonumber
\end{align}
Using the proposed identity (\ref{New_inde1}) we can convert the base of
quantum algorithm into
\begin{align}
U_{G}\left\vert 0\right\rangle ^{\otimes N}\left\vert v\right\rangle  &
=\left\langle \kappa\mathcal{G}_{N+1;0}\mathcal{G}_{1;N-1}\right\rangle
_{w}^{0,\left\vert +\right\rangle ^{\otimes N}\left\vert v\right\rangle
}K\left\vert 0\right\rangle ^{\otimes\left(  N+M\right)  }\label{EqUG1}\\
&  +\mu_{R_{0}\mathcal{G}_{N+1;0}\mathcal{G}_{1;N-1}}^{\left\vert
+\right\rangle ^{\otimes N}\left\vert v\right\rangle }\Delta_{\left\vert
+\right\rangle ^{\otimes N}\left\vert v\right\rangle }\left(  R_{0}%
\mathcal{G}_{N+1;0}\mathcal{G}_{1;N-1}\right)  \left\vert 0_{\perp
}\right\rangle .\nonumber
\end{align}
where $\kappa=\left\langle 0\right\vert ^{\otimes\left(  N+M\right)
}\left\vert +\right\rangle ^{\otimes N}\left\vert v\right\rangle /\left\langle
K\right\rangle ^{0},$ for a chosen quantum gate $K.$

\section{Discussion}

We have provided a general framework that allows the translation of quantum
algorithms into a quantum state in which the amplitudes are functions of the
expectation and standard deviation of the gates in the algorithm. Thus,
instead of operating each of the gates in the system, we can start with
constructing such quantum states to achieve the desired outcomes by measuring
such desired states. We further introduced a more fundamental version of the
AV identity that allows the translation of one quantum basis into another,
naturally giving rise to weak values. The construction of the desired quantum
state for a quantum algorithm with a sequence of gates $U_{G}=G_{M}%
G_{M-1}...G_{0}$ can be made by first computing the expectation and standard
deviation of each gate in the algorithm (or a sub-sequence of gates) and then
construct the states. The best-known upper bound for the number of gates that
allows us to construct the desired state is $O\left(  2^{N}\right)  $ for a
quantum algorithm that consists of $N$ qubits (see, [28]). The connection
between statistical measures and quantum algorithms is fundamental, following
the probabilistic nature of quantum mechanics, followed by the Born rule, and
the implicit quantum uncertainties. Thus, the proposed framework, which
combines statistical measures of each gate for obtaining a different
representation of the quantum algorithm, provides leads to a new understanding
of the relations between the quantum algorithms and the statistics of each of
the gates in the algorithm, which in our case, only relies on the two
statistical measures, the expectation and standard deviation of the gates. For
future research, we suggest investigating how the proposed framework can
optimize existing quantum algorithms. By focusing on the expectation and
standard deviation of the quantum gates, one could develop techniques to
minimize gate errors or maximize algorithmic efficiency, leading to more
stable quantum computations. Another avenue for future research is to explore
whether the proposed representation may improve quantum error correction
protocols, followed by the proposed partitions of translations of the quantum
gates (see, (\ref{sub_seq1}) with (\ref{Formula1})). There are, of course,
additional aspects of quantum algorithms that can be explored within the
proposed general framework.

\end{document}